\documentclass{jafm}
\usepackage{lipsum,booktabs}
\usepackage[caption=false]{subfig}
\usepackage{amsmath,amsfonts,amssymb}

\received_date{$-$}
\accepted_date{$-$}
\begin{document}
\title{Evaluation of machine learning algorithms for  predictive Reynolds stress transport modeling}
\author{
J P PANDA, H V WARRIOR}
\institute{Department of Ocean Engineering and Naval Architecture\\
              Indian Institute of Technology Kharagpur, India}

\email{jppanda.iit@gmail.com}
\abstract{The application machine learning (ML) algorithms to turbulence modeling has shown promise over the last few years, but their application has been restricted to eddy viscosity based closure approaches. In this article we discuss rationale for the application of machine learning with high-fidelity turbulence data to develop models at the level of Reynolds stress transport modeling. Based on these rationale we compare different machine learning algorithms to determine their efficacy and robustness at modeling the different transport processes in the Reynolds Stress Transport Equations. Those data driven algorithms include Random forests, gradient boosted trees and neural networks. The direct numerical simulation (DNS) data for flow in channels is used both as training and testing of the ML models. The optimal hyper-parameters of the ML algorithms are determined using Bayesian optimization. The efficacy of above mentioned algorithms is assessed in the modeling and prediction of the terms in the Reynolds stress transport equations. It was observed that all the three algorithms predict the turbulence parameters with acceptable level of accuracy. These ML models are then applied for prediction of the pressure strain correlation of flow cases that are different from the flows used for training, to assess their robustness and generalizability. This explores the assertion that ML based data driven turbulence models can overcome the modeling limitations associated with the traditional turbulence models and ML models trained with large amounts data with different classes of flows can predict flow field with reasonable accuracy for unknown flows with similar flow physics. In addition to this verification we carry out validation for the final ML models by assessing the importance of different input features for prediction.}
\nomenclature{
\begin{tabular}{l l}
$R_{ij}$ & Reynolds stress tensor\\
$b_{ij}$  & Reynolds stress anisotropy\\
$\delta_{ij}$  & Kronecker delta\\
$k$  & turbulent kinetic energy\\
$\rho$  & density\\
$\nu$  & Kinetic viscosity\\
$U_i$  & Mean velocity component\\
$u_i$ & Fluctuating velocity component\\
$P_{ij}$  & Production of turbulence\\
$T_{ijk}$  & Diffusive transport\\
$\epsilon_{ij}$  & Dissipation rate tensor\\
$\phi_{ij}$  & Pressure strain correlation\\
\end{tabular}
}
\maketitle
\section{Introduction}
\label{sec:intro}
Computational fluid dynamics(CFD) utilizes numerical techniques to model and solve problems that involve flow of fluids. CFD based methods\cite{versteeg2007introduction} have became popular with the advancement in computational facilities and formulation of models for turbulence\cite{pope2000}. The CFD based techniques are widely used in the early years of 21st century. The success of CFD techniques is largely dependent on the development of the techniques to model the turbulence. Mainly there are four broad classes of CFD approaches in the turbulence modeling framework, those are eddy viscosity models\cite{pope1975,pope2000,panda2020simple}, Reynolds stress transport models\cite{mishra2017toward,panda2018representation}, Large eddy simulation\cite{pope2000} and Direct numerical simulation\cite{lee2015direct,lee2018extreme}. In industrial applications eddy viscosity based are widely used, since those have least level of complexity and a few number of equations are solved for the modeling of the flow field. In recent years with booming in computational facilities researchers have started using LES for flow prediction in complex and larger domains. But LES can not be used for modeling and prediction complex flows of Engineering interest, since the cost of simulations of LES is very high. The LES and DNS data are mainly used for development and calibration of new turbulence models. In contrast to all the above mentioned approaches, the Reynolds stress transport models\cite{panda2018representation,mishra2017toward} are superior to eddy viscosity models, since in these models equations are solved for all the components of Reynolds stresses from which the Reynolds stress field is approximated. The building block of such models are the Reynolds stress transport equations\cite{panda2019review}, in which, the two important terms that need to be modeled are the pressure strain correlation\cite{panda2020reliable} and the dissipation term. The models for all these terms are developed by calibrating with few cases of experiments such as grid turbulence, grid turbulence with mean strain and shear flows. So, when these models are applied to other realistic cases of flow predictions may provide inaccurate results.

To overcome these difficulties recent emphasis of turbulence modelers have been shifted towards development of data-driven turbulence models\cite{duraisamy2019turbulence}. Machine learning is a process of function approximation, the function can be approximated by correlating the input parameters with the output. There are various machine learning algorithms. Few important of those are neural networks, Random forests\cite{wang2017physics,kaandorp2020data,luan2020influence}, gradient boosted trees\cite{wu2020comparative}, Gene expression programming\cite{zhao2020rans} and sparse symbolic regression\cite{schmelzer2020discovery}. In turbulence modeling, the problem is posed as one of supervised learning, where the model attempts to  minimize the prediction error over a training data-set. This data required for training, validation and testing the ML models can be obtained from experimental investigations, high fidelity DNS and LES data-sets of turbulent flows. In the context of supervised learning, data driven approaches of turbulence modelling can be classified into three main categories a) quantification of uncertainties in the RANS models (with this approach the uncertainties in Reynolds stress tensor can be quantified), b) finding the discrepancies in the model coefficients and magnitude of the terms in the governing equation and c) direct modeling of turbulence parameters such as Reynolds stress(EVM), pressure strain correlation (RSTM) and the subgrid scale stress(LES) in terms of the mean flow parameters. Among the three mentioned approaches of data-driven turbulence modeling, the models developed through the third approach can be trained with huge database of turbulence and can be applied for prediction of unseen flow cases of similar flow physics.

\cite{singh2017machine} developed model augmentations for Spalart-Allmaras(SA) turbulence model using adjoint based full field inference using experimentally measured lift coefficient data. These models forms are reconstructed using neural networks, and applied in CFD solver, to predict flow in different operating conditions. \cite{tracey2015machine} used a shallow neural network (one hidden layer) to model the source terms of the SA turbulence model. \cite{parish2016paradigm} learned a turbulence production term using machine learning and applied that to the $k$-equation of the $k-\omega$ turbulence model. \cite{maulik2018data} used neural networks to model eddy viscosities for RANS simulations. \cite{ling2016reynolds} used a special kind of neural network called tensor basis neural network to model the Reynolds stress anisotropy and also compared the model prediction against a simple multi layer perception. The optimal hyper-parameters of the model were recommended using Bayesian optimization. \cite{zhu2019machine} used  neural networks to construct a mapping function between the turbulent eddy viscosity and the mean flow variables and ML model completely replaces the partial differential equation model. \cite{wu2018physics} proposed a physics based implicit treatment to model Reynolds stresses using random forests. The optimal eddy viscosity and non-linear part of the Reynolds stresses were both predicted. They used seven distinct physics based features to model the Reynolds stresses, those are strain rate tensor, rotation rate tensor, pressure gradient, turbulence kinetic energy gradient, wall distance based Reynolds number, turbulence intensity and ratio of turbulent
time-scale to mean strain time-scale. They used different approaches to achieve in-variance in machine learning. \cite{weatheritt2016novel} used Gene expression programming to formulate non-linear consecutive stress-strain relationship. The mathematical model was created by using high fidelity and uncertainty measures. The learning method has the capability to produce a constraint free model. \cite{weatheritt2017development} used symbolic regression to model the algebraic form of the Reynolds stress anisotropy tensor. The equations were trained using hybrid RANS/LES data and the new model was employed in RANS closure to test the prediction of model in 3D geometries. \cite{schmelzer2020discovery} discovered algebraic Reynolds stress models using sparse regression and they have used high fidelity LES/DNS data for training and cross validation of the model. There case of separated flows were considered, those are periodic hills, converging-diverging channel and curved backward facing step. The prediction of the machine learnt model was better than the $k-omega$ SST model. \cite{huijing2021data} used the model developed by \cite{schmelzer2020discovery} to predict the fully three dimensional high Reynolds number flows, e.g. wall mounted cubes and cuboids. \cite{fang2020neural} used neural networks to model the Reynolds stress anisotropy using neural networks. They proposed different modification of the neural network structure to accommodate effect of Reynolds number, non-locality and wall effects into the modeling basis. With such distinct feature injection, significant improvement of model prediction was observed. \cite{beck2019deep} proposed a novel data-driven strategy for turbulence modeling for LES using artificial neural networks.

The methodology used in many of the studies discussed till now uses large high fidelity datasets obtained from DNS or LES simulations\cite{parashar2020modeling}. This data is used to train and validate machine learning based models, that can vary from deep learning based models or ensembled meta-models. The training of this model involves the inference of optimal coefficients for the closure of the turbulence model. In these studies, the turbulence model form corresponds to 2-equation eddy viscosity based models (EVM) or Algebraic Reynolds Stress Models (ARSM). While these studies have had considerable success, this outlined methodology may be hamstrung by the discordance between the high fidelity of datasets and the fidelity of the baseline model form. As an illustration DNS data is able to replicate the high degree of anisotropy in turbulent flows. However eddy viscosity based models are not capable of replicating a high degree of turbulence anisotropy due to the linear eddy viscosity hypothesis\cite{mishra2019estimating}. The eddy viscosity hypothesis assumes that turbulence anisotropy is a linear function of the instantaneous mean strain rate. Thus the anisotropy predicted by any eddy viscosity based model is forced to lie on the plane strain segment on the barycentric triangle \cite{edeling2018data}. As a result the information regarding the turbulence anisotropy in the high fidelity data is ignored by the machine learning algorithm, due to the model form of the baseline eddy viscosity closure. Additionally the high fidelity data reflects the complicated dependence of turbulent statistics on streamline curvature and the mean rate of rotation. This is important information to model the mean and frame rotation effects on turbulence evolution. However eddy viscosity based closures model the Reynolds stresses as functions of  the mean rate of strain only\cite{pope2001turbulent}. Thus the information regarding the effect of streamline curvature or frame rotation is ineffectual because of the form of the baseline model \cite{mishra2019linear}. Algebraic Reynolds Stress Models assume that  turbulent diffusive and convective fluxes are relatively small (explicitly that the flow is source dominated \cite{gatski1993explicit}). This is a restrictive assumption and is not valid for most turbulent flows.  In light of these illustrations the adoption of a different baseline model formulation, flexible enough to take advantage of the high fidelity information in the data may be advisable. An alternative that can meet these requirements is the Reynolds Stress Modeling approach.  Instead of assuming the form of relationship between the Reynolds stresses and the mean gradient, the Reynolds stress models use the Reynolds Stress Transport Equations as evolution equations for components of turbulent anisotropy. Explicit computation of the evolution of components of turbulent anisotropy results in improved representation of anisotropy. Reynolds stress models can account for directional effects of the Reynolds stresses. Reynolds stress models can represent the limiting states of turbulent flow, for example the return to isotropy of turbulence in decaying turbulent flows, and the  Rapid Distortion Limit where turbulence behaves like an elastic medium (instead of a viscous medium). Because of the separate modeling of the turbulent transport processes, Reynolds stress models account for effects of flow stratification, buoyancy, streamline curvature, etc. Thus in a machine learning framework, utilizing Reynolds Stress Models as the baseline model can enable the algorithm to exploit a substantially higher proportion of physics and information in the high fidelity data. 

However there has been little research to extend the potential of  Reynolds Stress Modeling approach by utilizing machine learning algorithms. This is a central novelty of this investigation. 

In the Reynolds Stress Modeling approach separate models are formulated for the terms in the Reynolds Stress Transport Equation, where each such term represents a different turbulence transport process. These transport processes include turbulent diffusion, rotational effects, rate of dissipation and the pressure strain correlation. While high fidelity models for all these terms are important, accurate and robust modeling of the pressure strain correlation term is a long standing challenge in turbulence modeling. The pressure strain correlation term represents physics responsible for the transfer of energy between different components of the Reynolds stress tensor \cite{mishra2014realizability}. It is responsible for the non-local interactions in turbulent flows, the initiation of instabilities in rotation dominated flows, the return to isotropy observed in decaying flows, etc \cite{mishra2013intercomponent}. While classical models have been developed for the pressure strain correlation term, such physics driven models have many limitations in their ability to account for streamline curvature effects, realizability requirements, their performance in complex engineering flows \cite{mishra2017toward}. In this work we have modeled the pressure strain correlation of turbulence using three different machine learning approaches, those are artificial neural network, random forest and gradient boosted decision trees. The ML models developed with above mentioned algorithms are trained and tested for DNS data of turbulent channel flow at different Reynolds numbers. We have grouped the data-sets in 4 different combination to perform 4 distinct training and testing of the ML models. The Bayesian hyper-parameter optimization was also used to find the best hyper-parameters of the ML models that perform well for unknown prediction problems. One main advantage of hyper-parameter optimization is that it reduces the chance of over-fitting in tree based algorithms and also enhance the generalizability of the model.           

The remainder of the article is summarized as follows: Section 2 provides a background on the turbulence modeling framework. In  Section 3 the limitations of the turbulence models in predicting complex flow fields is discussed. Section 4 describes the different machine learning algorithms used in modeling the turbulence parameters. In section 5 the detailed methods of optimization of the hyper-parameters of the ML models are discussed. In the subsequent sections the training-testing of the models and feature importance are discussed followed by conclusion and future scope.           
\section{Reynolds stress transport modeling framework}
\label{sec:rastm framework}
Different fidelities of turbulence simulation approaches can be differentiated based on the scales of motion that are directly simulated based on their governing equations and the scales of motion that are modeled. For example, in Direct Numerical Simulation, all scales of motion are computed. In Reynolds Averaged Navier Stokes (RANS) approaches, all turbulent scales of motion are modeled. RANS closures can carry out this modeling of turbulent scales of motion in different manner. One equation models may use the concept of mixing length to model turbulent scales of motion. Two-equation models like the $k-\epsilon$ and the $k-\omega$ models use the eddy viscosity hypothesis to model turbulent motions. Reynolds Stress Models are based off the Reynolds Stress Transport Equations. Here, each turbulent transport process is present as a separate term in the evolution equations. In Reynolds stress transport modeling, such transport terms such as pressure strain correlation and dissipation tensors can be modeled using data driven approaches. In this section we discuss the traditional method of modeling of the pressure strain correlation starting from the basic governing equations.

The Reynolds stress transport model\cite{mishra1,panda2018representation,panda2017} does not rely upon any ad hoc definitions  of turbulent stresses in terms of strain field, rather the stress field is directly computed from the modeled transport equation of Reynolds stress components. The Reynolds stress transport equation has four terms in its right side, those are production, transport, dissipation and the pressure strain correlation term. The pressure strain correlation term mainly accounts for the complex flow features resulting from the stream line curvature and flow separation. The Reynolds stress model much more reliable and accurate than its two equation counter parts.
The Reynolds stress transport equation has the form\cite{panda2019review}:
\begin{equation}
\begin{aligned}
&\partial_{t} \overline{u_iu_j}+U_k \frac{\partial \overline{u_iu_j}}{\partial x_k}=P_{ij}-\frac{\partial T_{ijk}}{\partial x_k}-\epsilon_{ij}+\phi_{ij},\nonumber\\ &
\mbox{where},\nonumber\\ & 
P_{ij}=-\overline{u_ku_j}\frac{\partial U_i}{\partial x_k}-\overline{u_iu_k}\frac{\partial U_j}{\partial x_k}, \nonumber\\ & T_{kij}=\overline{u_iu_ju_k}-\nu \frac{\partial \overline{u_iu_j}}{\partial{x_k}}+\delta_{jk}\overline{ u_i \frac{p}{\rho}}+\delta_{ik}\overline{ u_j \frac{p}{\rho}}\nonumber\\ &
,\epsilon_{ij}=-2\nu\overline{\frac{\partial u_i}{\partial x_k}\frac{\partial u_j}{\partial x_k}},\nonumber\\ &\phi_{ij}= \overline{\frac{p}{\rho}(\frac{\partial u_i}{\partial x_j}+\frac{\partial u_j}{\partial x_i})}
\end{aligned}
\end{equation}
$P_{ij}$ denotes the production of turbulence, $T_{ijk}$ is the diffusive transport, $\epsilon_{ij}$ is the dissipation rate tensor and $\phi_{ij}$ is the pressure strain correlation. The pressure fluctuations are governed by a Poisson equation:
\begin{equation}
\frac{1}{\rho}{\nabla}^2(p)=-2\frac{\partial{U}_j}{\partial{x}_i}\frac{\partial{u}_i}{\partial{x}_j}-\frac{\partial^2 u_iu_j}{\partial x_i \partial x_j}
\end{equation}
The fluctuating pressure term is split into a slow and rapid pressure term $p=p^S+p^R$. Slow and rapid pressure fluctuations satisfy the following equations
\begin{equation}
\frac{1}{\rho}{\nabla}^2(p^S)=-\frac{\partial^2}{\partial x_i \partial x_j}{(u_iu_j-\overline {u_iu_j})}
\end{equation}
\begin{equation}
\frac{1}{\rho}{\nabla}^2(p^R)=-2\frac{\partial{U}_j}{\partial{x}_i}\frac{\partial{u}_i}{\partial{x}_j}
\end{equation}
It can be seen that the slow pressure term accounts for the non-linear interactions in the fluctuating velocity field and the rapid pressure term accounts for the linear interactions\cite{mishra4}. The pressure strain correlation is modeled using rational mechanics approach. The rapid term can be modeled as\cite{pope2000}
\begin{equation}
\phi_{ij}^R=4k\frac{\partial{U}_l}{\partial{x_k}}(M_{kjil}+M_{ikjl})
\end{equation}
where, 
\begin{equation}
M_{ijpq}=\frac{-1}{8\pi k}\int \frac{1}{r} \frac {\partial^2 R_{ij}(r)}{\partial r_p \partial r_p}dr
\end{equation}
where, $R_{ij}(r)=\langle u_i(x)u_j(x+r) \rangle$
For homogeneous turbulence the complete pressure strain correlation can be written as
\begin{equation}
\phi_{ij}=\epsilon A_{ij}(b)+kM_{ijkl}(b)\frac{\partial\overline {v}_k}{\partial{x_l}}
\end{equation}
The most general form of slow pressure strain correlation is given by
\begin{equation}
\phi^{S}_{ij}=\beta_1 b_{ij} + \beta_2 (b_{ik}b_{kj}- \frac{1}{3}II_b \delta_{ij})
\end{equation}
Established slow pressure strain correlation models including the models of \cite{rotta1951} and \cite{ssmodel,panda2017,warrior2014} use this general expression. Considering the rapid pressure strain correlation, the linear form of the model expression is
\begin{equation}
\begin{aligned}
\frac{\phi^{R}_{ij}}{k}=C_2 S_{ij} +C_3 (b_{ik}S_{jk}+b_{jk}S_{ik}-\\\frac{2}{3}b_{mn}S_{mn}\delta_{ij})+  
C_4 (b_{ik}W_{jk} + b_{jk}W_{ik})
\end{aligned}
\end{equation}
Here $b_{ij}=\frac{\overline{u_iu_j}}{2k}-\frac{\delta_{ij}}{3}$ is the Reynolds stress anisotropy tensor, $S_{ij}$ is the mean rate of strain and $W_{ij}$ is the mean rate of rotation. Rapid pressure strain correlation models like the models of \cite{mishra2017toward,speziale1991modelling,panda2018representation} use this general expression. For this work the Reynolds stress model of \cite{speziale1991modelling} is used which has the form:

\begin{equation}
\begin{aligned}
& \phi_{ij}^{(R)}=(C_1-C_1^*II^{0.5})K S_{ij}+\\ & C_2K(b_{ik} S_{jk}+b_{jk} S_{ik}-2/3b_{mn} S_{mn}\delta_{ij})\\ & +C_3K(b_{ik} W_{jk}+b_{jk} W_{ik})
\end{aligned}
\end{equation}
\begin{figure}
\centering
 \subfloat[]{\includegraphics[width=0.45\textwidth]{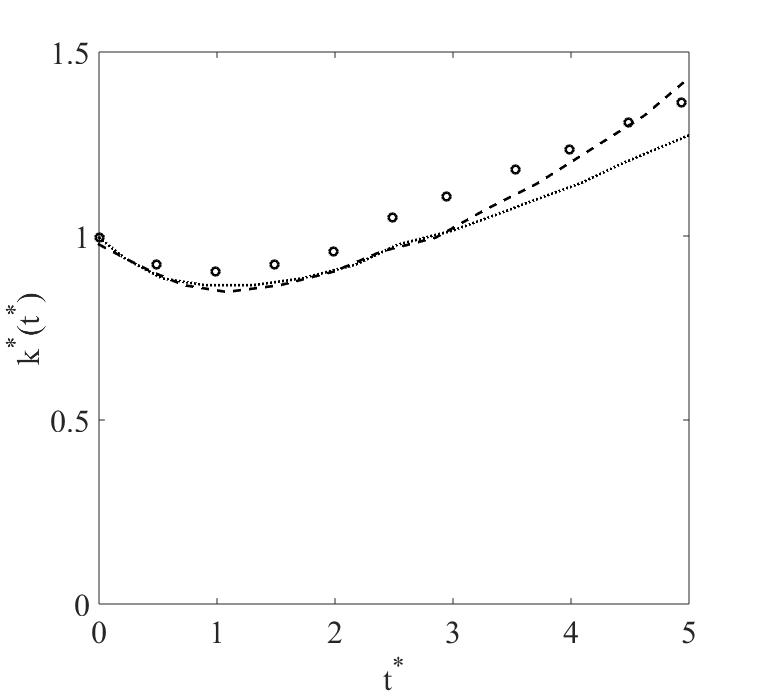}}\\
 \subfloat[]{\includegraphics[width=0.45\textwidth]{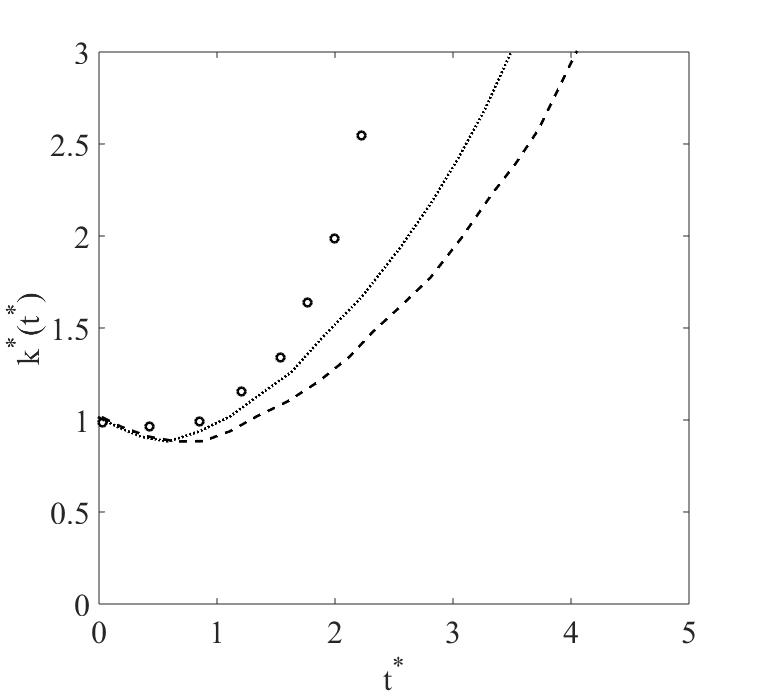}}\\
 \subfloat[]{\includegraphics[width=0.45\textwidth]{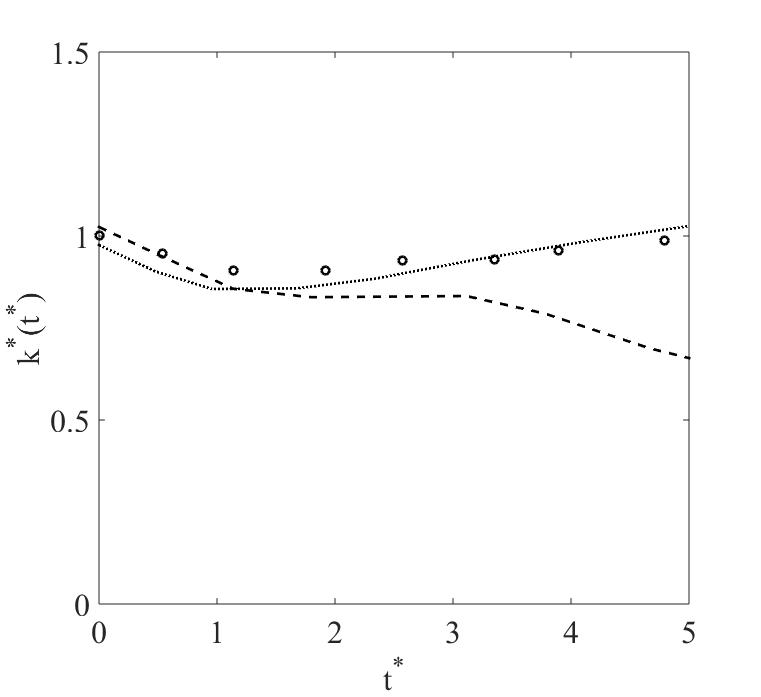}}
 \caption{Evolution of turbulence kinetic energy for rotating shear flows, a) $W/S=0$, b) $W/S=0.25$ and c) $W/S=0.5$, Symbols LES data, dashed lines predictions of LRR model, dotted lines predictions of SSG model.\label{fig:1}}
\end{figure}
The closure coefficients are taken as  $C_{1}=0.8$, $C_1^{*}=1.3$, $C_2=1.25$ and $C_3=0.4$.

\section{Limitation of the existing pressure strain correlation model in the prediction of complex turbulent flows}
\label{sec:limitations}
Most of the Reynolds stress transport models available in literature have some tuned coefficients, those are obtained from the calibration of set of equations against few standard experimental/DNS data-sets like flow in channels, grid turbulence and grid turbulence, so when those models are tested for complex flow configurations produce unrealistic results. The Reynolds stress transport models are also associated with another type of uncertainty, i.e. those can not accurately capture the non-local nature of flow(The flow at one point may be affected by flow physics of upstream points), since the model form does not have any terms for accommodation of the non-local nature of flow. In order to overcome these limitations of the Reynolds stress transport models data driven models of the pressure strain correlation model can be developed using machine learning approaches.         

We have performed numerical simulations to check the predictive capability of different pressure strain correlation models in rotation dominated flow fields. It is noticed that, there is a strong disparity between the model predictions and the LES data particularly at higher values of rotation to strain ratio. The results of those comparisons are presented in fig.\ref{fig:1}. fig.\ref{fig:1} a, b and c correspond to rotation to strain ratio of $0$, $0.25$ and $0.5$ respectively. The symbols in the fig.\ref{fig:1} represent the LES results of\cite{bardino1983improved}.      
\section{Description of the data set}
The dataset used in the modeling of the pressure strain correlation of turbulence was obtained from the Oden Institute Turbulence File Server\cite{lee2015direct}, in which DNS simulations were performed for flow in channels with friction Reynolds number ranging from 550 to 5200. The DNS simulations were performed with a B-spline collocation method in the wall-normal direction and for the stream-wise and span-wise directions Fourier-Galerkin method was used. The detailed information on the on the methods employed in the DNS of channel flow is available in \cite{lee2015direct}. We have used four distinct data-sets for training and testing of the ML models. The four training and testing data-sets are presented in table\ref{t2}.   

\begin{table}
\begin{center}
 \begin{tabular}
 {||c  c c||} 
 \hline
 Case & Training Set & Testing Set \\ [0.5ex] 
 \hline\hline
 1 & $Re_\lambda=550,1000,2000$  & $Re_\lambda=5200$ \\ 
 \hline
 2 & $Re_\lambda=550,1000,5200$ &  $Re_\lambda=2000$ \\
 \hline
 3 & $Re_\lambda=550,2000,5200$ &  $Re_\lambda=1000$ \\
 
 \hline
 4 & $Re_\lambda=1000,2000,5200$ & $Re_\lambda=550$ \\ [1ex] 
\hline
\end{tabular}
\end{center}
\caption{Four training and test cases for the turbulent channel flow\label{t2}}
\end{table}
\section{Physics based input features for ML models}
The turbulence parameter e.g. pressure strain correlation can be accurately modeled using suitable input features. The most important features for the modeling of the pressure strain correlation can be the Reynolds stress anisotropy, dissipation, velocity gradient and the turbulence kinetic energy. The input features must be wisely chosen so that there is no over-fitting. Secondly this ensures that physics based constraints are met in the final model. For instance due to Galilean in-variance we should ensure that the features in the modeling basis also obey this requirement. The functional mapping for the ML models can be written as: 
\begin{equation}
\begin{aligned}
&\phi_{uv}=f_1(b_{uv}, \epsilon, \frac{du}{dy}, k)
\end{aligned}
\end{equation}
We have used the formula: $\alpha^*=\frac{\alpha-\alpha_{min}}{\alpha_{max}-\alpha_{min}}$ for normalizing the inputs to the ML models, so that those will be in the range 0 and 1. This avoids clustering of training in one direction and enhance convergence in the training.  
\section{Data driven turbulence modeling with machine learning}
\label{sec: Data driven turbulence}
\begin{figure*}
\centering
\subfloat[]{\includegraphics[height=3.6cm]{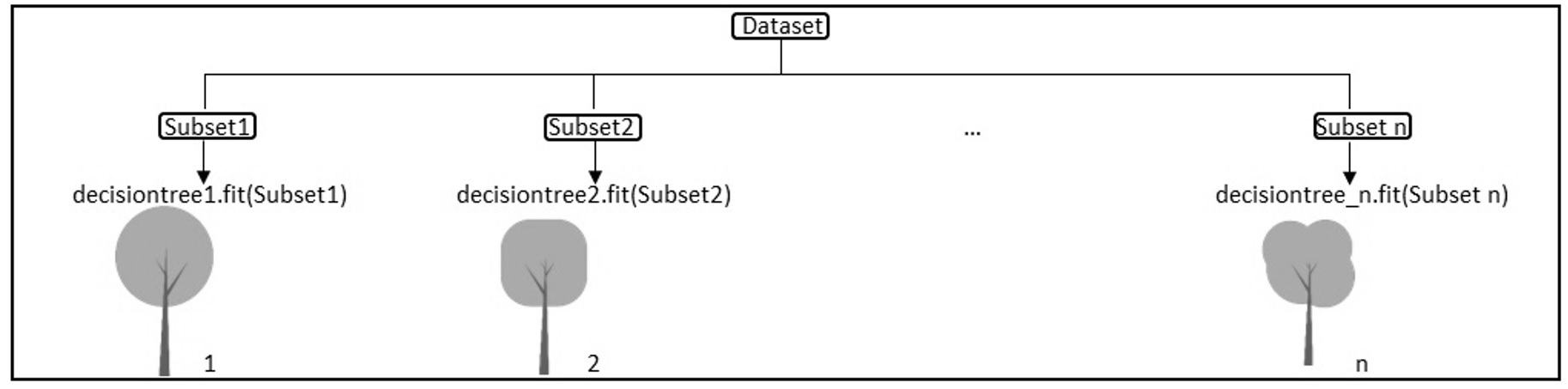}}\\
\subfloat[]{\includegraphics[height=3.6cm]{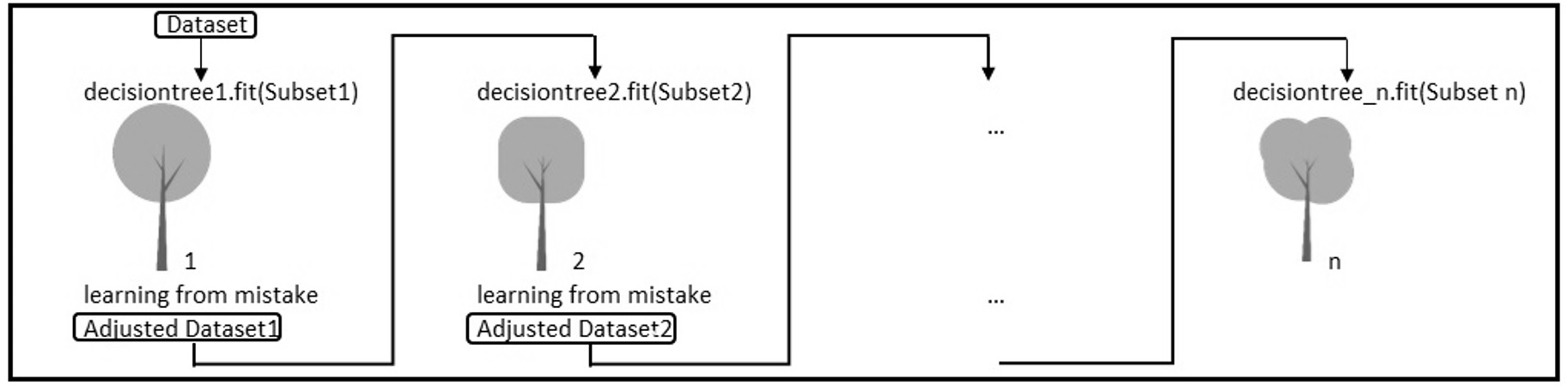}}
\caption{a) Bagging vs b) Boosting} \label{fig:2}
\end{figure*}

In turbulence modeling frameworks such as eddy viscosity and Reynolds stress transport models the Reynolds stress and pressure strain correlation terms plays major role and with accurate modeling of those two parameters, the model form uncertainties can be eradicated. In this work, we have modeled above mentioned terms using three different machine learning algorithms, those are deep neural networks, Random forests and gradient boosted trees. In turbulence modeling and fluid dynamics applications deep neural networks are widely used. Next, we have used random forests in the models, those have the capability to check the feature importance, i.e. which of the input features has dominant correlation with the output. We also have considered gradient boosted trees in the modeling of the turbulence parameters. The basic difference between RF and GB trees is that RF creates random training samples from full training set based on Bootstrap Aggregating(Bagging) and a weak learner is trained in parallel and in GB trees ensemble methods consist in fitting several weak learners sequentially as shown in figure. \ref{fig:2}.   

\subsection{Artificial Neural Networks}
\begin{figure*}
\centering
\subfloat[]{\includegraphics[height=6cm]{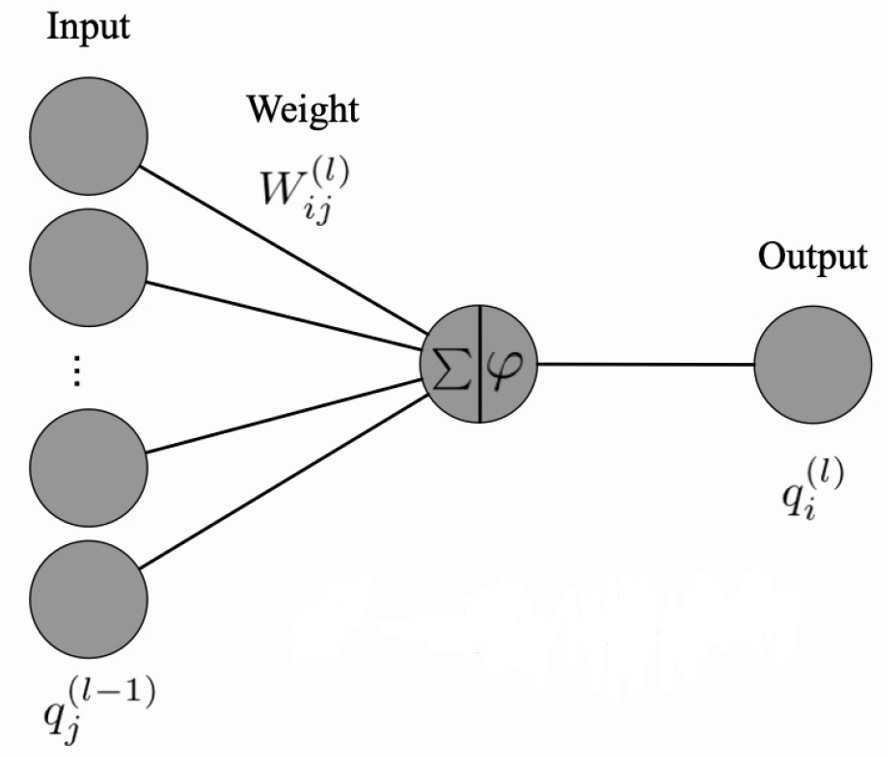}}
\subfloat[]{\includegraphics[height=8cm]{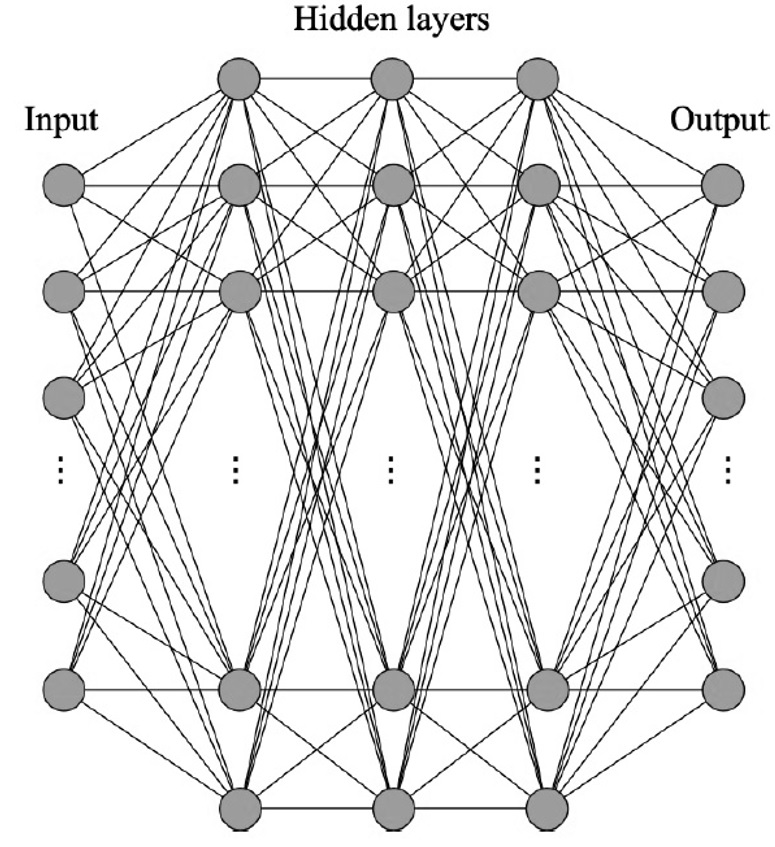}}
\caption{a) A perceptron b) An artificial neural network(Multi layer perception)} \label{fig:3}
\end{figure*}

Artificial Neural Networks(ANN)(fig.\ref{fig:3}) are the machine learning systems inspired from the biological neural networks, these are also known as multi-layer perception(MLP). The biological neural networks(BNN) are the circuits that carry out a specific task when activated. These are population of neurons interconnected by synapses. Similar to BNNs, ANN/MLP have artificial neurons. A very basic unit of a MLP is a perceptron as shown in fig.\ref{fig:3}a. The input data from a (l-1)th layer are multiplied by a weight $W$, which are linearly combined and allowed to pass through a non-linear activation function $\eta$: 
\begin{equation}
\begin{aligned}
& q_{i}^{l}=\eta(\sum_j W_{ij}^{l}q_{j}^{l-1}).
\end{aligned}
\end{equation}
A MLP can be constructed by combining a number of perceptrons. A typical MLP is shown in fig.\ref{fig:3}b. The weights in a MLP are optimized to minimize a cost function $E$ by back propagation. There are several optimization techniques by which the weights can be calculated. Those are gradient descent, Quasi- Newton, Stochastic Gradient Descent or the Adaptive moment Estimation etc. The activation functions, which can be used in MLP are sigmoid $\eta(\beta)=1/(1+e^{-\beta})$, hyperbolic tangent(tanh) $\eta(\beta)=(e^{-\beta}-e^{-\beta}).(e^{-\beta}+e^{-\beta})$ and RELU $\eta(\beta)=max[0,\beta]$.
\subsection{Random forests(RF)}
\begin{figure*}
\centering
\includegraphics[height=6.2cm]{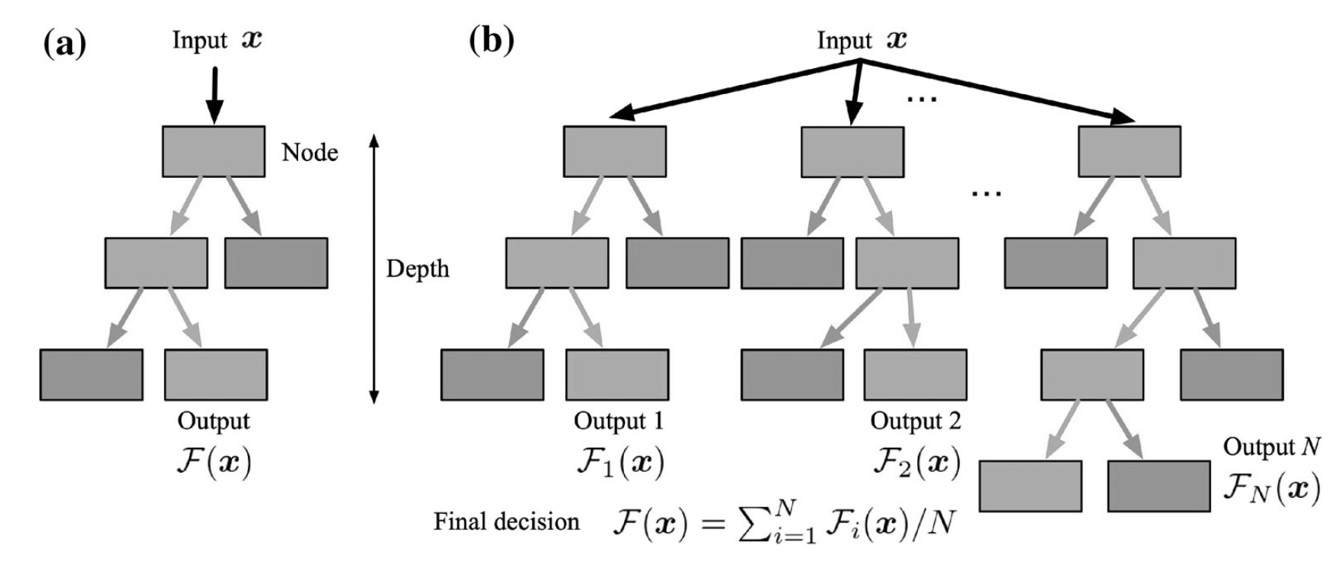}\\
\caption{Architecture of of a random forest, a)a single decision tree b)a random forest} \label{fig:4}
\end{figure*}

The random forest algorithm was proposed by \cite{breiman2001random}. The basic building block of a random forest is a decision tree. The structure of the decision tree is shown in fig.\ref{fig:4}a. The boxes in the decision tree represent group of features and data. The decision tree seeks if/then/else rules for obtaining the desired output. Random forest(RF)(fig.\ref{fig:4}b) regression combines performance of multiple decision trees for predicting a output variable. It is a assemble learning technique, works on the concept of bagging method. In RF regression trees are constructed using a subset of random sample drawn with replacement from training data. In RF to depict the growth of the tress random vectors are generated and the trees are not allowed to prun. To perform splitting of the dataset, a random combination of features is selected at each and every node.      

\subsection{Gradient boosted decision trees(GBDT)}
Similar to random forest algorithm, the building block of GBDT are the decision trees. Random forests utilizes method of bagging to combine many decision trees to create an ensemble. Bagging simply means combining the decision trees in parallel. However, boosting means combining decision trees in series to achieve strong learner. The decision trees are the weak learners. The boosting algorithms learn slowly, since trees are added sequentially. Each tree in the boosting trees focus on errors from previous one, making the boosting algorithm an efficient and accurate model. Boosting process is slow, since the trees in the GBDT are added sequentially.  
\section{Hyper-parameter optimisation}
\label{sec:Hyper-parameter}
The hyper-parameter optimization is the process of finding the best hyper-parameters for a machine learning algorithm that returns best performance while measured on a validation set. The hyper-parameters are turned by application engineers during model developments. There are mainly two methods of hyper-parameter tuning, those are either performed manually or done by some optimization algorithm. The automated algorithms are grid search, random search and Bayesian optimization. The grid search and random search not informed by past evaluations. The Bayesian optimization keep track of past evaluations, hence we have considered the Bayesian optimization for automated hyper-parameter tuning.       
\subsection{Manual search}
The hyper-parameters of different machine learning algorithms, which need to finely tuned for improving the predictive capability of any machine learning model. For neural networks the important hyper-parameters are number of layers and number of neurons in each layer. Similarly for random forest and gradient boosted decision trees the hyper-parameters are number of decision trees and maximum depth of the trees. Few other parameters are also important in tree algorithms, those are minimum sample leaf and learning rate for gradient boosting decision trees. However fine tuning of all such parameters using manual search method is tedious task and also there is also the chance of over fitting. In manual search, we have considered $R^2$ as the criteria for finding the hyper-parameters. By manual search the optimal hyper-parameters are tuned as follows: for neural network: 5 layers with 10 neurons in each layer and for both RF and GBDT the number of decision trees and max depth are taken as 5 and 10 respectively. By considering above mentioned hyper-parameters the $R^2$ and MSE values for test data are calculated as follows for case 4: ANN: $0.9897$, $9.97*10^{-6}$, RF: $0.9841$, $1.66*10^{-5}$, GBDT: $0.981$, $1.99*10^{-5}$.            
\subsection{Bayesian optimization}
In Bayesian optimization, the parameter space is sampled and a gradient boosted tree was constructed and trained. The performance of the model is evaluated using validation data. The optimal hyper-parameters are those, that yield the lowest error on the validation data. The model validation error was treated as a sample from Gaussian process at each hyper-parameter setting. In contrast to other optimization techniques, the Bayesian optimization requires relatively lesser number of model evaluations, since at each step the information from all previous states is utilized to inform the GP model of the validation error. This method is highly useful, when computational cost of evaluation of objective function is very high. The detailed steps in Bayesian optimization are as follows:\\~\\
1. Build a surrogate probability model of the objective function\\
2. Finding the hyper-parameters that performs best on the surrogate model\\
3. Application of these hyper-parameters to the true objective function\\
4. Updating of the surrogate model incorporating the new results\\
5. The steps 2 and 4 must be repeated until max iterations or time is reached\\~\\

\begin{figure}
\centering
\subfloat[]{\includegraphics[height=6cm]{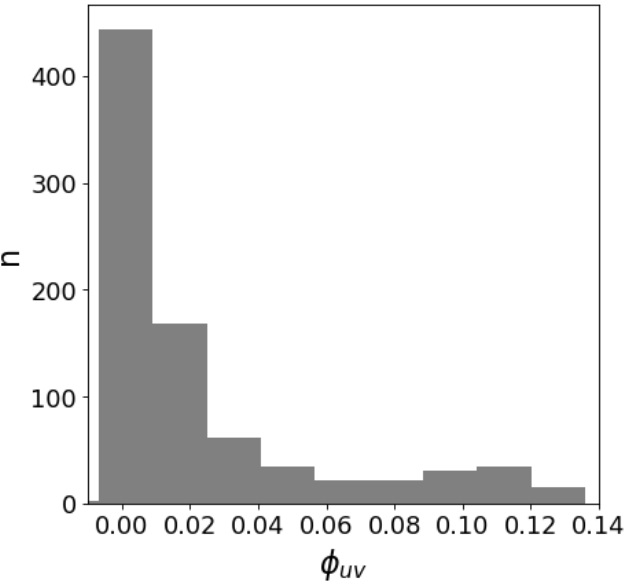}}
\caption{Histogram of the data variability for training data in case 1, There are very few training samples for high values of pressure strain correlation.}\label{fig:hist}
\end{figure}
\begin{table}
\begin{center}
 \begin{tabular}
 {||c c c ||} 
 \hline
 ML models & Parameter I & parameter II\\ [0.5ex] 
 \hline\hline
 MLP & N. of neurons & Layers \\ 
 \hline
 RF & N. of decision trees  & Depth\\
 \hline
 GBDT & N. of decision trees  & Depth\\
 
\hline
\end{tabular}
\end{center}
\caption{The hyper-parameters of the machine learning models. (N. stands for number).\label{t1}}
\end{table}
\begin{figure*}
\centering
 \subfloat[]{\includegraphics[width=0.5\textwidth]{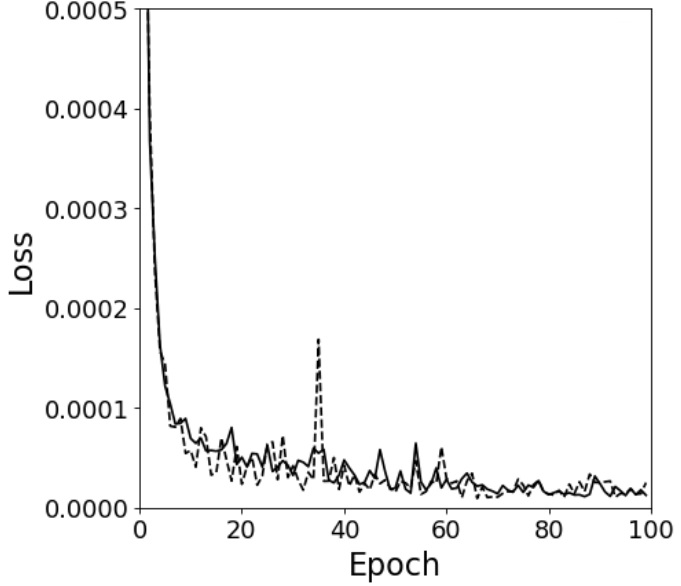}}
 \subfloat[]{\includegraphics[width=0.5\textwidth]{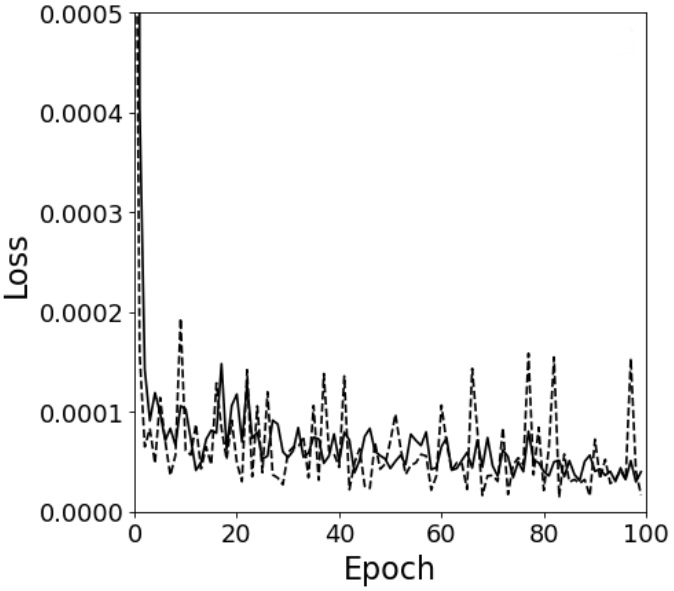}}\\
 \subfloat[]{\includegraphics[width=0.5\textwidth]{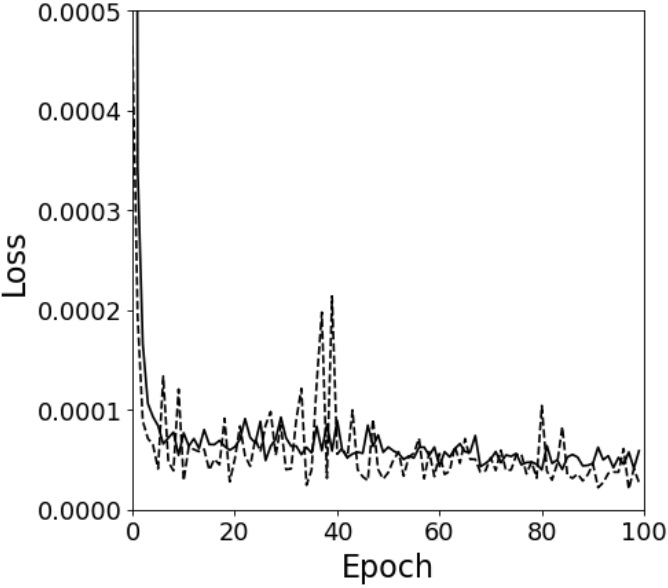}}
  \subfloat[]{\includegraphics[width=0.5\textwidth]{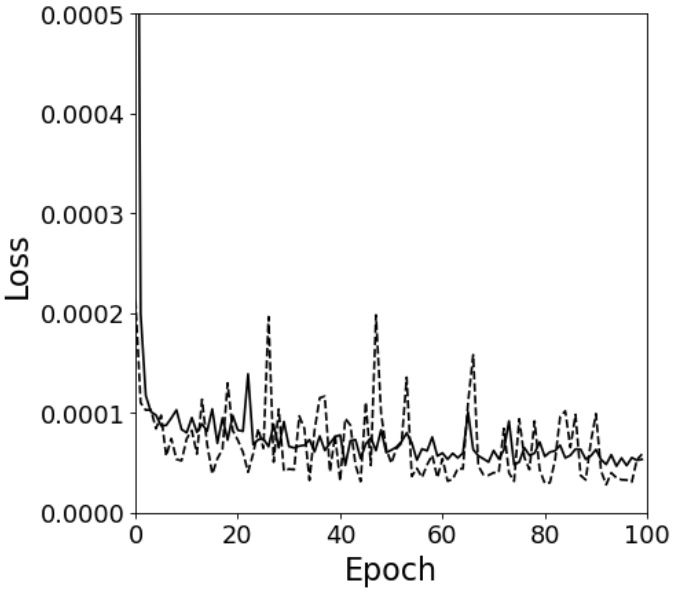}}
 \caption{Loss convergence in MLP training. a) case 4 b) case 3 c) case 2 d) case 1. Solid line and dashed line correspond to training and validation losses respectively.\label{fig:5}}
\end{figure*}
\begin{table*}[h]
\begin{center}
\begin{tabular} {c c c c c c c c c} 
  \toprule
  \bfseries  &  \multicolumn{2}{c}{\bfseries Case 1}  &  \multicolumn{2}{c}{\bfseries Case 2} &  \multicolumn{2}{c}{\bfseries Case 3}  &  \multicolumn{2}{c}{\bfseries Case 4} \\
  \cmidrule(lr){1-9}
  \bfseries Models &  \multicolumn{1}{c}{\bfseries Train $R^2$} & \multicolumn{1}{c}{\bfseries Test $R^2$} & \multicolumn{1}{c}{\bfseries Train $R^2$} & \multicolumn{1}{c}{\bfseries Test $R^2$}  &  \multicolumn{1}{c}{\bfseries Train $R^2$} & \multicolumn{1}{c}{\bfseries Test $R^2$} & \multicolumn{1}{c}{\bfseries Train $R^2$} & \multicolumn{1}{c}{\bfseries Test $R^2$}\\
  \midrule
  MLP & 0.967 & 0.978 & 0.978  & 0.984 & 0.921 & 0.891 & 0.992  & 0.892 \\
  RF & 0.971 & 0.978 & 0.949  & 0.966 & 0.962 & 0.988 & 0.984  & 0.885 \\
  GBDT & 0.982  & 0.871  & 0.968  & 0.958 & 0.951 & 0.955 & 0.982  & 0.871 \\
\bottomrule
\end{tabular}
\caption{$R^2$ of $\phi_{uv}^*$ predictions by training-prediction cases}\label{t3}
\end{center}
\end{table*}
The hyper-parameters mentioned in table \ref{t1} are only a limited set of important parameters, through which the model can be developed. The models with those limited tuned parameters may not perform well for unknown prediction data-sets on which the model is trained. To enhance the predictive capability of any machine learning model at least few other hypo-parameters must be considered in the model development stage. However, with increase in number of hyper-parameters, manual tuning may be a tedious process and advanced optimization techniques like Bayesian optimisation must be used for fine turning of the hyper-parameters.   
For brevity, we have only considered the hyper-parameter optimization of gradient boosted regression tree using Bayesian optimization. In addition to depth and number of decision trees, three other hypo-parameters are considered, those are max features, minimum simple leaf and minimum sample split. From Bayesian optimization, the optimal hyper-parameters(maximum depth, Number of estimators, minimum sample split, minimum sample leaf, maximum features) are (66, 63, 9, 5, sqrt). The maximum depth correspond to highest depth of the individual regression tree. The best value of this parameter depends on the interaction of the input variables. Number of estimators are the number of boosting stages used in the model development. The larger is the number of the estimators, better is the performance of the the model. The minimum sample leaf is the minimum number of minimum number of samples required to split an internal node. The other terms has their usual meaning. A detailed discussion of all these hyper-parameters of GBDT is available in \cite{scikit-learn}. The $R^2$ values of the optimized GBDT was found to be 0.9915 and 0.9076 for the training and testing respectively for case 4.  The predictions of the GBDT with optimized hyper-parameters for unknown flow cases are summarized in section \ref{sec:testing}.    
\section{Training of the ML models}
\label{sec:training}
\begin{figure*}
\centering
 \subfloat[]{\includegraphics[width=0.5\textwidth]{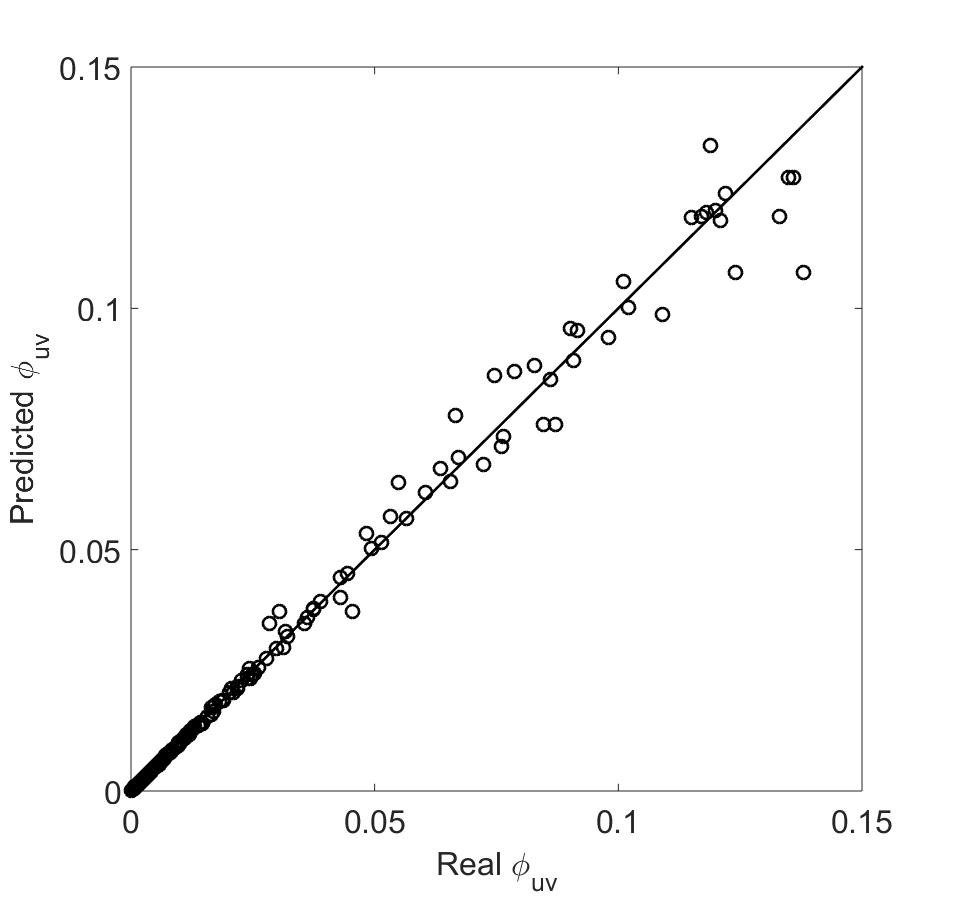}}
\subfloat[]{\includegraphics[width=0.5\textwidth]{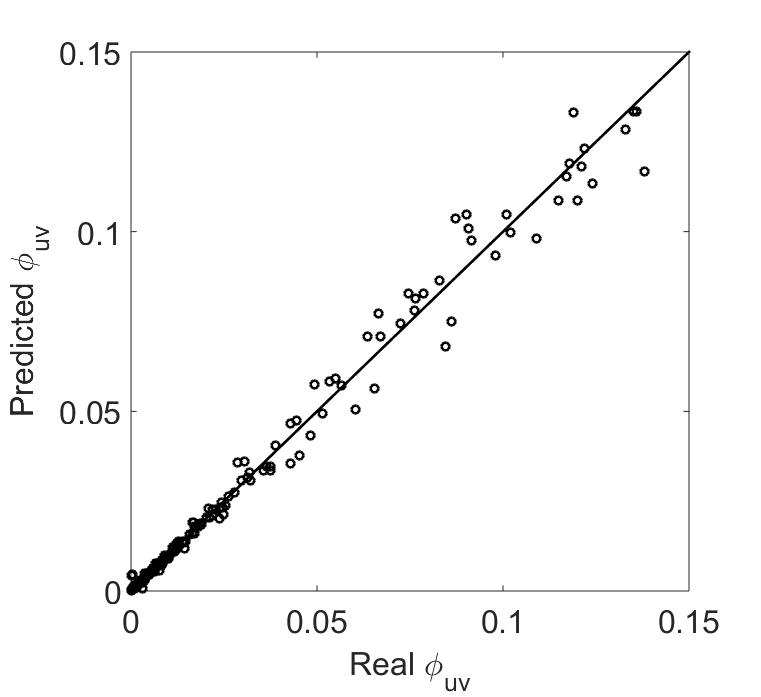}}\\
\subfloat[]{\includegraphics[width=0.5\textwidth]{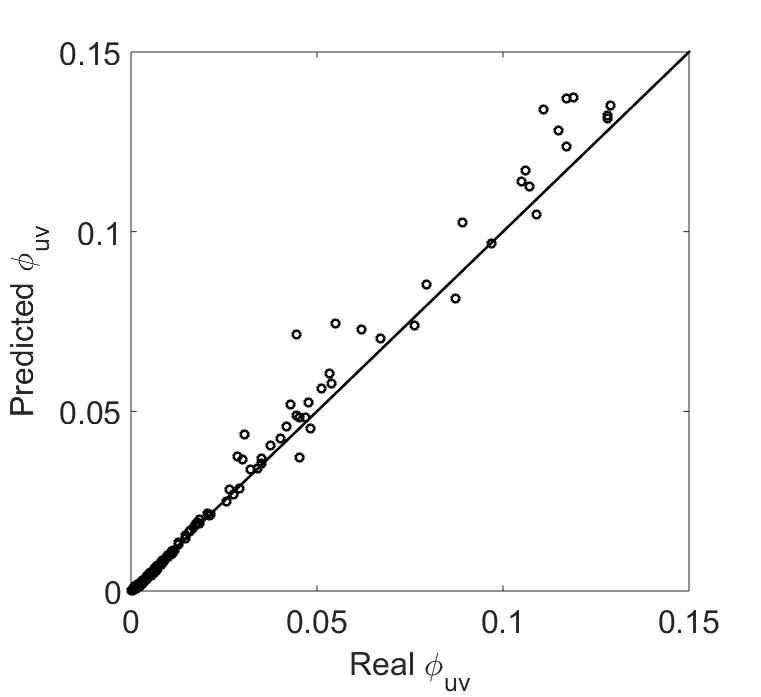}}
 \caption{DNS simulation samples versus data driven model predictions of the pressure strain correlation for the ML models training for case 4, a) Random forest b) Gradient boosted trees c) Neural network.\label{fig:6}}
\end{figure*}

All the three ML models are trained for the DNS data of turbulent channel flow for four different test cases. The four different cases are shown in table \ref{t2}. Since the turbulence statistics are available for four different friction Reynolds numbers, in each training and testing phase, we have considered data for three friction Reynolds numbers as training and the other was considered for testing. We have considered two different approaches for finding the hyper-parameters of the ML models as discussed in section 6. Using the hyper-parameters obtained from manual search, the $R^2$ scores obtained from different ML models are presented in table \ref{t3}. For case 1, the ML models were trained with data for friction Reynolds numbers 550, 1000, 2000 and tested for 5200. The maximum value of train $R^2$ is obtained from the results of case 4 with MLP. The loss convergence history of MLP for the four different cases are shown in figure.\ref{fig:5}. In most of the training cases, the $R^2$ magnitude is greater than $0.95$, which signifies a good correlation between the actual and predicted value of pressure strain term ($\phi$) for training samples (A $R^2$ value of 0 and 1 correspond to no-correlation and perfect correlation between the actual and predicted values of the pressure strain term). In figure.\ref{fig:6}, the results of DNS simulation samples versus data driven model predictions of the pressure strain correlation are presented. There is very good match between the DNS results and DNS model predictions. There are very few points in the data-set with high values of the pressure strain correlation, this signifies the fact that, the relative inaccuracy of the model predictions for high values of the pressure strain correlation is due to very few such training data samples(figure\ref{fig:hist}). 
\begin{figure*}
\centering
\subfloat[]{\includegraphics[width=0.5\textwidth]{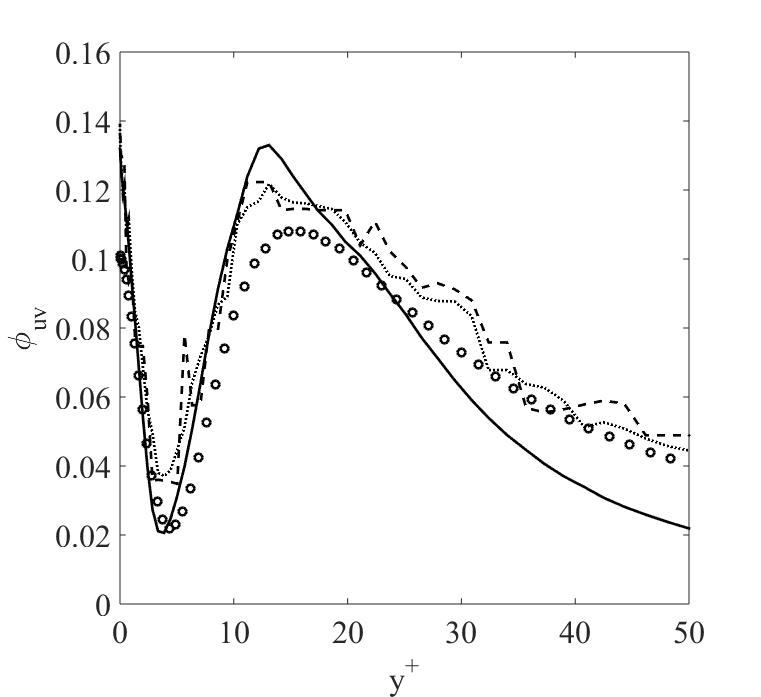}}
\subfloat[]{\includegraphics[width=0.5\textwidth]{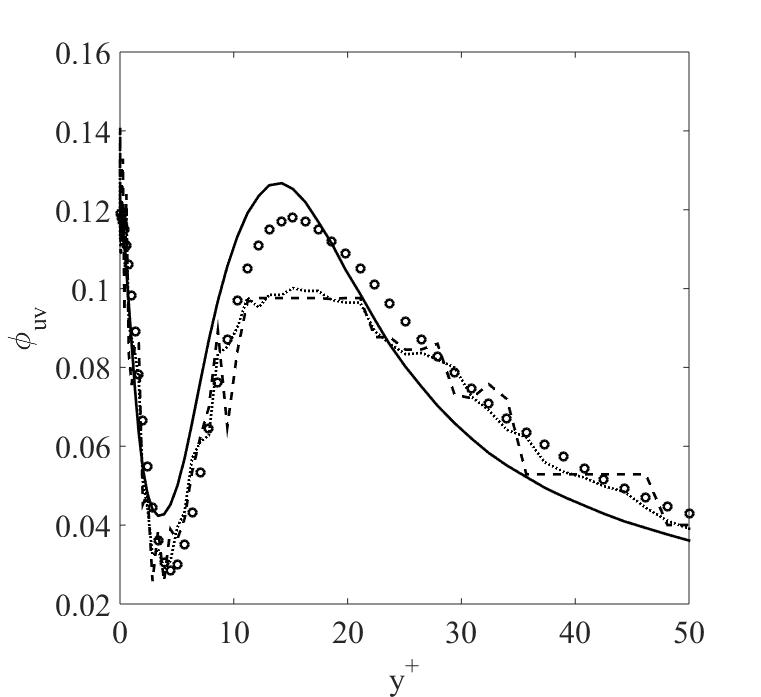}}\\
\subfloat[]{\includegraphics[width=0.5\textwidth]{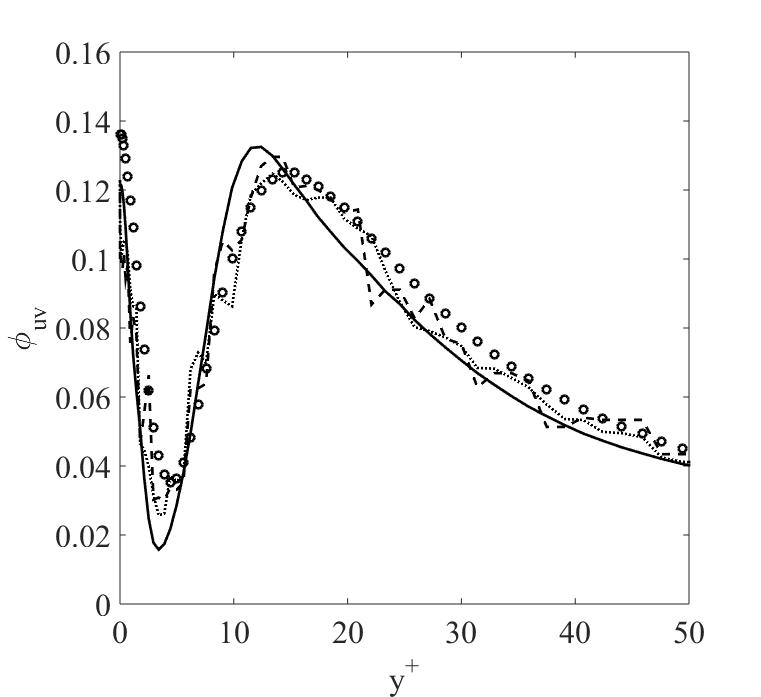}}
\subfloat[]{\includegraphics[width=0.5\textwidth]{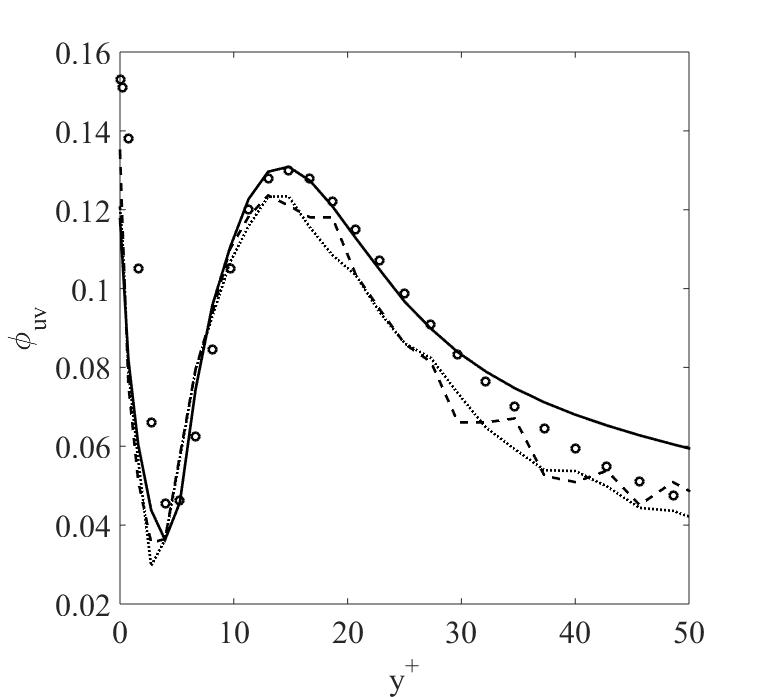}}
 \caption{Prediction of the pressure strain correlation using different machine learning algorithms.Solid line, dashed line and dotted line represent predictions of neural network, gradient boosted trees and random forests respectively. a) case 4, b) case 3, c) case 2, d) case 1.\label{fig:7}}
\end{figure*}
\section{Testing of the trained ML models}
\label{sec:testing}
\begin{figure}
\centering
 \subfloat[]{\includegraphics[width=0.5\textwidth]{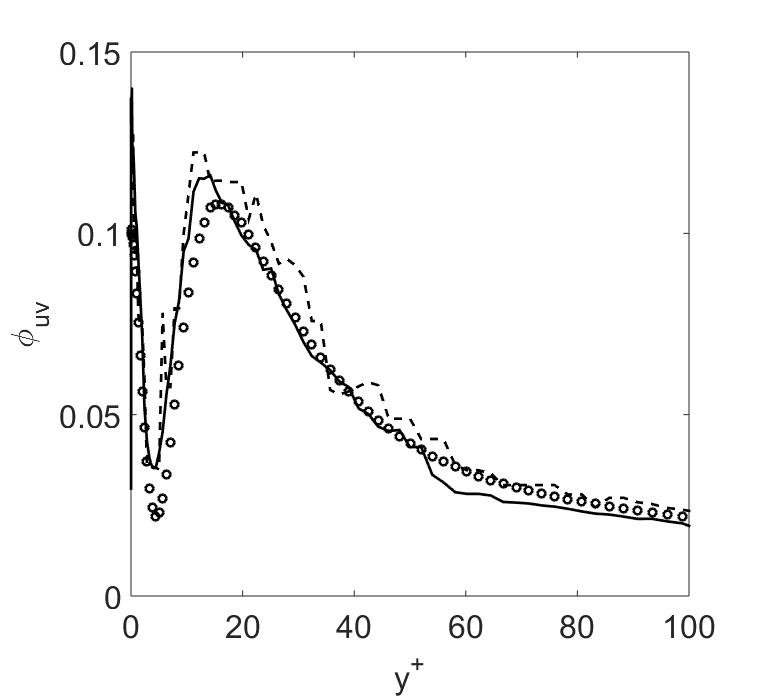}}
 \caption{Prediction of the pressure strain correlation for case 4, using the optimized GBDT. The hyper-parameters of the GBDT were tuned using Bayesian optimization.\label{fig:8}}
\end{figure}

\begin{figure}
\centering
 \subfloat[]{\includegraphics[width=0.5\textwidth]{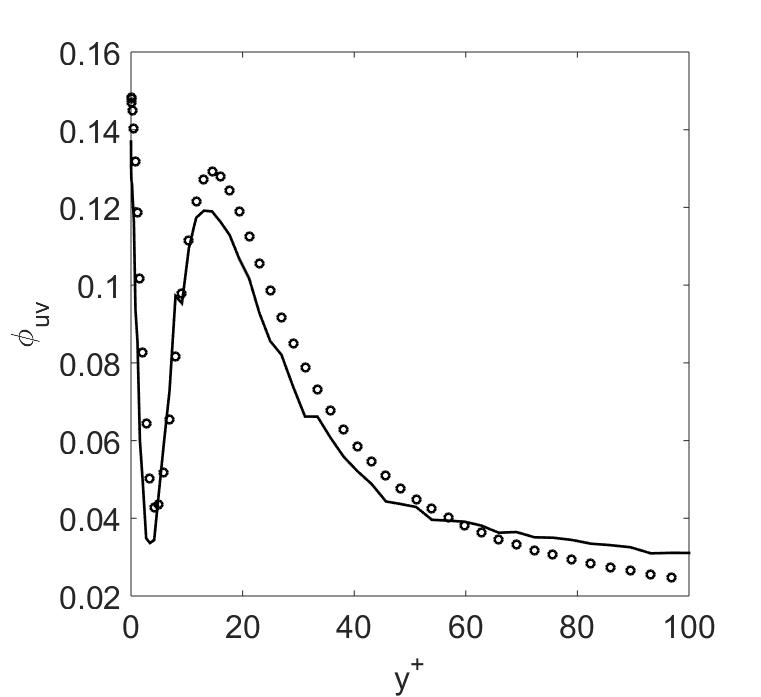}}
 \caption{Prediction of the pressure strain correlation for turbulent Couette flow at $Re_\lambda=550$.\label{fig:9}}
\end{figure}

The trained models were tested against testing data for all the four cases as discussed in table.\ref{t2}. The results of testing of the ML models are presented in \ref{fig:7}. From all four figures of figure \ref{fig:7} it is clear that, the GBDT predictions are better in comparison to the predictions of MLP and RF. The neural networks failed to predict the pressure strain correlation for $y+$ values greater than 25 for case 4. The testing results are presented in table \ref{t3}. In most of the testing cases the $R^2$ value was found to be greater than $0.95$, but in few cases like case 4, the $R^2$ value falls below $0.9$ for the testing case.  
This is because of over-fitting of the models with the training data. The over-fitting problem is common in random forests. The advantage of GBDT over MLP is that the former needs less number of weights in developing the correlation between the inputs and the output. Since the ultimate aim of turbulence modeler is to apply the ML based based into CFD solver, models with less number of weights are always preferred. Since model with large number of weights, may end with divergence in CFD solver. So we have tried to enhance the predictive capability of the GBDT with Bayesian optimization of hyper-parameters as discussed in section \ref{sec:Hyper-parameter}. The results of predicted values $\phi$ with the optimized GBDT is presented in \ref{fig:8}. The solid line in the figure correspond to the predictions of the optimized GBDT and the dashed line correspond the prediction of the GBDT with the hyper-parameters obtained from manual tuning. from the figure it is clear that, the predictions of the optimized GBDT matches well with the DNS results. Finally, we have tested the GBDT model against a fully unknown flow e.g. Couette flow in channels. The model predictions are presented in figure.\ref{fig:9}. It is noticed that the GBDT predictions are matching well the DNS results of \cite{lee2018extreme}.        

\section{Conclusions}
In this article, we discuss rationale for the application of machine learning with high-fidelity turbulence data to develop models at the level of Reynolds stress transport modeling. Then, we ascertain the efficacy of different machine learning algorithms at creating surrogate models for the pressure strain correlation. We have modeled the pressure strain correlation of turbulence using three different machine learning approaches, those are Random forests, gradient boosted tress and artificial neural networks. The input features to the ML models were chosen from the traditional modeling basis of the pressure strain correlation, those are mean strain, turbulence kinetic energy, Reynolds stress anisotropy and turbulence dissipation. The ML models were trained and tested for DNS data of turbulent channel flow at different friction Reynolds numbers. The optimal values of ML model hyper-parameters were optimized using manual search and Bayesian optimization approaches. The feature importances of different input features of the random forest were obtained using mean decrease in impurity method. It was noticed that the mean strain has larger correlation with the pressure strain term. The ML models developed by using data-driven approaches can be utilized in computational fluid dynamics solvers for improved flow predictions in turbulent channel flows. in future course of work, for achieving universality, large amount of DNS or experimental data can be fed to the ML models during training and development. 
\bibliography{jafm}
\end{document}